\documentclass[graybox]{svmult}

\usepackage{mathptmx}       
\usepackage{helvet}         
\usepackage{courier}        
\usepackage{type1cm}        
\usepackage{makeidx}         
\usepackage{graphicx}        
\usepackage{multicol}        
\usepackage[bottom]{footmisc}

\makeindex             

\begin{document}

\title*{Science with the EXTraS Project: Exploring the X-ray Transient and variable Sky.}
\titlerunning{The EXTraS Project.} 
\author{A. De Luca, R. Salvaterra, A. Tiengo, D. D'Agostino, M.G. Watson,
F. Haberl, J. Wilms \emph{on behalf of the EXTraS collaboration}.}
\authorrunning{A. De Luca et al.} 
\institute{
A. De Luca \at INAF-IASF Milano, Via Bassini 15, I-20133 Milano, Italy \email{deluca@iasf-milano.inaf.it}
\and R. Salvaterra \at INAF-IASF Milano, Via Bassini 15, I-20133 Milano, Italy \email{ruben@iasf-milano.inaf.it}
\and A. Tiengo \at IUSS Pavia, Piazza della Vittoria 15, I-27100 Pavia, Italy \email{andrea.tiengo@iusspavia.it}
\and D. D'Agostino \at CNR-IMATI, Via de Marini 6, I-16149 Genova, Italy \email{dagostino@ge.imati.cnr.it}
\and M.G. Watson \at University of Leicester, Department of Physics and Astronomy
Leicester, LE1 7RH, UK\\ \email{mgw@leicester.ac.uk}
\and F. Haberl \at MPG-MPE, Giessenbachstrasse, D-85748 Garching, Germany \email{fwh@mpe.mpg.de}
\and J. Wilms \at ECAP, Sternwartstrasse 7, D-96049 Bamberg, Germany\\ \email{joern.wilms@sternwarte.uni-erlangen.de}
}
%
%
\maketitle

\abstract*{The EXTraS project (``Exploring the X-ray Transient
and variable Sky'') will characterise the temporal
behaviour of the largest ever sample of objects in the soft X-ray range 
(0.1-12 keV) with a complex, systematic and consistent
analysis of all data collected by the European Photon
Imaging Camera (EPIC) instrument onboard the ESA XMM-Newton
X-ray observatory since its launch. 
We will search for, and characterize variability
(both periodic and aperiodic) in hundreds of thousands
of sources spanning more than nine orders of magnitude
in time scale and six orders of magnitude in flux.
We will also search for fast transients, missed by
standard image analysis. Our analysis will be completed by 
multiwavelength characterization of new discoveries
and phenomenological classification of variable sources. 
All results and products will be made available to the community
in a public archive, serving as a reference for a broad
range of astrophysical investigations.
}

\abstract{The EXTraS project (``Exploring the X-ray Transient
and variable Sky'') will characterise the temporal
behaviour of the largest ever sample of objects in the soft X-ray range 
(0.1-12 keV) with a complex, systematic and consistent
analysis of all data collected by the European Photon
Imaging Camera (EPIC) instrument onboard the ESA XMM-Newton
X-ray observatory since its launch. 
We will search for, and characterize variability
(both periodic and aperiodic) in hundreds of thousands
of sources spanning more than nine orders of magnitude
in time scale and six orders of magnitude in flux.
We will also search for fast transients, missed by
standard image analysis. Our analysis will be completed by 
multiwavelength characterization of new discoveries
and phenomenological classification of variable sources. 
All results and products will be made available to the community
in a public archive, serving as a reference for a broad
range of astrophysical investigations.}
\section{The EXTraS project: aim, implementation.}
\label{sec:1}
The EXTraS project (``Exploring the X-ray Transient and variable Sky'')
will systematically explore the temporal domain
information stored in the database collected by the European Photon Imaging Camera
(EPIC, \cite{strueder01,turner01}) onboard the ESA XMM-Newton observatory. 
EPIC is the most powerful tool to study faint
X-ray sources in the 0.1-12 keV range. Indeed, the Serendipitous Source
Catalogue based on EPIC data, 
listing more than 560,000 detections in its most recent release (3XMM, see \cite{rosen15}), 
is the largest catalogue of X-ray sources ever compiled. However, time-domain 
information in such data, although very rich, remained mostly unused.  \\
EXTraS will release to the community a full temporal characterization
(both aperiodic and periodic variability)
of hundreds of thousand of sources with flux spanning from 
$10^{-9}$ to $10^{-15}$ erg cm$^{-2}$ s$^{-1}$ (0.2-10 keV),
on time scales ranging from $\sim0.1$s to $\sim10$ years.
EXTraS will also search for short-duration transients,
not included in 3XMM,
and will also perform a phenomenological classification of all sources
based on their temporal and spectral properties.
The different lines of temporal analysis   
implemented within the project 
are described
in more detail in \cite{deluca15}. \\ 
The project has received funding within the EU-FP7 framework 
(grant agreement n. 607452) and is carried out by a collaboration
including INAF (Italy), IUSS (Italy), CNR/IMATI (Italy),
University of Leicester (UK), MPE (Germany) and ECAP
(Germany).
More information, updates on the project
as well as a full list of contacts can be found by visiting
the project web site at \url{www.extras-fp7.eu}.

\section{An (incomplete) overview of Science with EXTraS.}
\label{sec:2}
\begin{figure}[h]
\includegraphics[width=\textwidth]{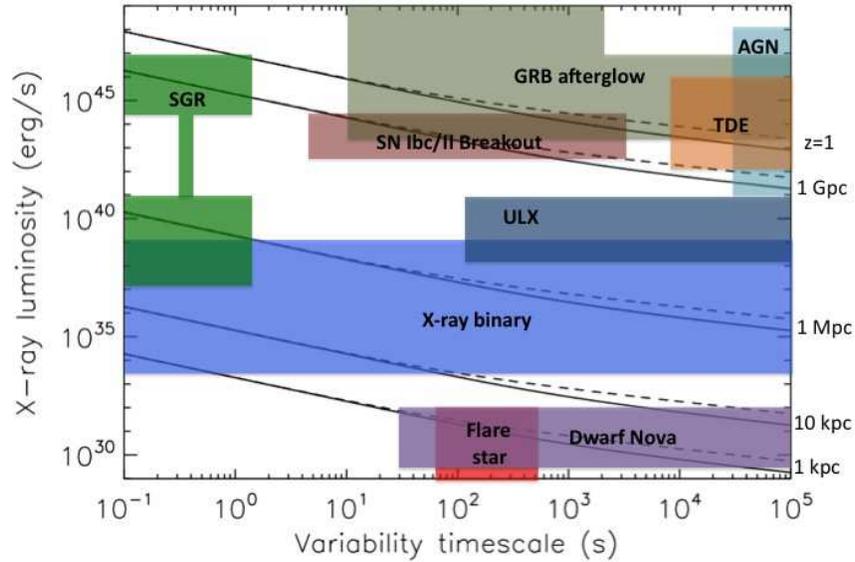}
%
%
\caption{Peak X-ray luminosity of different source populations as a function of their
characteristic variability time scale (adapted from \cite{merloni12}). 
Diagonal lines mark the approximate EXTraS 0.5-2 keV sensitivities (at $4\sigma$ confidence,
EPIC/pn camera only) for various source distances. 
Solid/dashed lines correspond to min/max particle background levels.
Even assuming a very conservative $6.5\sigma$ sensitivity level
(a factor 2--5 higher flux than in the figure),
implying an essentially null contamination from spurious detections,
we can expect a very rich harvest of variable sources, based
on the equivalent sky coverage of EPIC data and on previous observational 
results and theoretical models. Among such sources, several hundreds of
flares from a broad population of stars and protostars; 3-300 and 0.3-30 bursts 
from persistent and transient LMXBs, respectively, 
in the Galactic Center region; 3-5 SFXTs
active in bright flaring and 40-75 sources in an intermediate 
state of flaring; several new magnetars; a few TDEs up to
z$\sim5-6$; up to a dozen low-luminosity GRBs up to z$\sim2$; about 
ten SN shock break-out events up to z$\sim0.1$;
moreover, it can be estimated that about 100 ``Fast Radio Burst'' should have
occurred within the field of an ongoing EPIC pointed observation.}
\label{fig:1}       
\end{figure}

The extremely broad range of variability timescales and luminosities investigated 
by EXTraS is shown in Figure.~\ref{fig:1}. The scientific discovery space is very large 
and we trust that EXTraS results will have a great impact in many areas of astrophysics 
and cosmology. The blind nature 
of our search will allow astronomers to measure (or give strong limits on) the intrinsic 
occurrence rate of different transient events and to perform population studies. 
Withouth demanding completeness,
we include below a list of science cases that will benefit of our results (see
also caption to Fig.~\ref{fig:1}).\\
{\bf Flaring stars}: constraining, on a statistical basis,
the duration, duty cycle and amount of
energy released in flares of different stellar groups. Indeed, 
as an interesting first result of EXTraS, we may mention 
the unexpected detection of a flare
from a \emph{very young protostar} (Pizzocaro et al., submitted to A\&A)\\
{\bf Cataclysmic variables and Novae}: unveiling periodicities and bursts 
in about 100 known sources; searching for new systems. \\
{\bf Low-mass X-ray binaries (LMXBs)}:  probing e.g. the properties 
of bursting LMXBs \cite{strohmayer06} --
in particular, for the poorly known ``burst only sources''.
Bursts from M31, the closest spiral galaxy, can also be detected\cite{pietsch05}.\\
{\bf High-Mass X-ray Binaries (HMXBs)}: constraining e.g. the census 
of Supergiant Fast X-ray Transients (SFXT) \cite{sidoli11},
which is crucial to unveil the evolutionary path and
formation rate of massive stars.\\
{\bf Isolated Neutron Stars}: constraining e.g. 
the population of magnetars (Anomalous X-ray Pulsars, AXPs, and Soft Gamma Repeaters, SGRs) \cite{mereghetti08} 
and its relation to the overall population of Neutron Stars (NS), which will also impact
on our comprehension of the short GRBs.
The case of GRB150301C/3XMM J004514.7+415035, 
with properties consistent with those of an active magnetar in M31, are a clear
demonstration of the potentialities of EXTraS in this field (see GCN\,17548 and ATel\,\#7181).\\
{\bf ULtraluminous X-ray Sources (ULXs)}: assessing accretion physics
in these poorly understood sources \cite{feng11}
by e.g. searching for, and unveiling, source ``states''; systematically searching
for bursts and periodicities (orbital, or rotational)\\
{\bf Tidal Disruption Events (TDEs)}: constraining
their poorly known statistics 
(even the case of no detections would be interesting) 
and their physics \cite{burrows11}.\\
{\bf Gamma-Ray Bursts (GRBs)}: constraining the rate of Low-Luminosity GRBs 
(e.g. \cite{campana06}), possibly up to high redshift, and their progenitors;
no detections would set the strongest available constraint on their population.\\
{\bf Supernovae (SNe)}: inferring a measure of the SN rate
independent from optical surveys
by detecting SN shock breakout events \cite{soderberg08}; no detections would set 
important constraints on the poorly known underlying physics. \\
{\bf Active Galactic Nuclei (AGN)}: measuring the mass of 
$\sim100$ AGNs, based on their variability properties;
calibrating the variability-luminosity relation locally;
characterizing the variability of Blazars, their
duty-cycle on short timescales and deriving constraints on
particle injection/acceleration, magnetic field and BH mass.\\
{\bf Rare events}: ranging from galactic events 
(e.g. TDEs of minor bodies falling onto NSs or BHs) 
to cosmic ones (e.g. orphan afterglows of GRB seen off-axis, GRBs from the first
very massive stars); constraining
the high energy properties of the puzzling ``Fast Radio Bursts'' \cite{thornton13}.\\
{\bf Totally unexpected discoveries} can be also foreseen, 
as has always been the case when a new region in parameter space has been explored.

%

\end{document}